\documentclass[10pt,twocolumn]{article}
\usepackage{graphicx}

\begin{document}

\title{Relativistic mechanism of superconductivity}
\author{{\small H. Y. Cui} \\
{\small Department of Applied Physics}\\
{\small Beijing University of Aeronautics and Astronautics}\\
{\small Beijing, 100083, China}}
\date{{\small \today}}
\maketitle

\begin{abstract}
{\small According to the theory of relativity, the relativistic Coulomb's
force between an electron pair is composed of two parts, the main part is
repulsive, while the rest part can be attractive in certain situations. Thus
the relativistic attraction of an electron pair provides an insight into the
mechanism of superconductivity. In superconductor, there are, probably at
least, two kinds of collective motions which can eliminate the repulsion
between two electrons and let the attraction being dominant , the first is
the combination of lattice and electron gas, accounting for traditional
superconductivity; the second is the electron gas themselves, accounting for
high $T_c$ superconductivity. In usual materials, there is a good balance
between the repulsion and attraction of an electron pair, the electrons are
regarded as free electrons so that Fermi gas theory plays very well. But in
some materials, when the repulsion dominates electron pairs, the electron
gas will has a behavior opposite to superconductivity. In the present paper
the superconducting states are discussed in terms of relativistic quantum
theory in details, some significant results are obtained including quantized
magnetic flux, London equation, Meissner effect and Josephson effect. 
\newline
\newline
}
\end{abstract}

\section{Introduction}

In BCS theory, it is believed that the mechanism responsible for the
transition to superconductivity is a coupling between electrons via the
positive ions of metallic lattice. The electron-lattice-electron interaction
provides an attraction between electrons which can lead to a ground state
separated from excited states by an energy gap. Whereas, in recent years the
discovery of high $T_c$ superconductivity has offered a challenge for BCS
theory, the first great difficult in the extension of BCS theory is to
discover a nature of interaction responsible both for the traditional and
high $T_c$ superconductivities.

In the present paper, we propose a mechanism for superconductivity which is
based on the relativity theory.

\section{Relativistic Coulomb's force}

This section is a theoretical preparation for the next section.

Consider a particle moving in a inertial system with 4-vector velocity $u$,
it satisfies\cite{Harris}

\begin{equation}
u_\mu u_\mu =-c^2  \label{1}
\end{equation}
The above equation is valid so that any force can never change $u$ in its
magnitude but can change $u$ in its direction. We therefore conclude that
the relativistic Coulomb's force on a particle always acts in the direction
orthogonal to the 4-vector velocity of the particle in the 4-dimensional
space-time, rather than along the line joining a couple of particles.
Simply, any 4-vector force $f$ satisfy the following orthogonal relation

\begin{equation}
u_\mu f_\mu =u_\mu m\frac{du_\mu }{d\tau }=\frac m2\frac{d(u_\mu u_\mu )}{%
d\tau }=0  \label{2}
\end{equation}

\begin{figure}[htb]
\includegraphics[bb=175 580 355 740,clip]{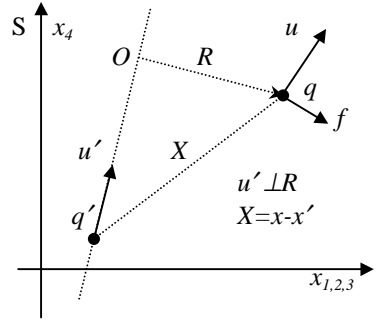}
\caption{The Coulomb's force acting on $q$ is perpendicular to the 4-vector
velocity $u$ of $q$, and lies in the plane of $u^{\prime }$ and $X$ with the
retardation with respect to $q^{\prime }$.}
\label{sfig1}
\end{figure}

Suppose there are two charged particle $q$ and $q^{\prime }$ locating at
positions $x$ and $x^{\prime }$ in the Cartesian coordinate system $S$ and
moving at 4-vector velocities $u$ and $u^{\prime }$ respectively, as shown
in Fig.\ref{sfig1}, where we use $X$ to denote $x-x^{\prime }$. The
Coulomb's force $f$ acting on particle $q$ is perpendicular (orthogonal) to
the velocity direction of $q$, as illustrated in Fig.\ref{sfig1}, like a
centripetal force, the force $f$ should make an attempt to rotate itself
about its path center, the center may locate at the front or back of the
particle $q^{\prime }$, so the force $f$ should lie in the plane of $%
u^{\prime }$ and $X$, then

\begin{equation}
f=Au^{\prime }+BX  \label{e1}
\end{equation}
Where $A$ and $B$ are unknown coefficients, the possibility of this
expansion was discussed in details in the paper\cite{Cui}, in where the
expansion is not an assumption. Using the relation $f\perp u$, we get

\begin{equation}
u\cdot f=A(u\cdot u^{\prime })+B(u\cdot X)=0  \label{e2}
\end{equation}
we rewrite Eq.(\ref{e1}) as

\begin{equation}
f=\frac A{u\cdot X}[(u\cdot X)u^{\prime }-(u\cdot u^{\prime })X]  \label{e3}
\end{equation}
It follows from the direction of Eq.(\ref{e3}) that the unit vector of the
Coulomb's force direction is given by

\begin{equation}
\widehat{f}=\frac 1{c^2r}[(u\cdot X)u^{\prime }-(u\cdot u^{\prime })X]
\label{e4}
\end{equation}
because

\begin{eqnarray}
\widehat{f} &=&\frac 1{c^2r}[(u\cdot X)u^{\prime }-(u\cdot u^{\prime })X] 
\nonumber \\
&=&\frac 1{c^2r}[(u\cdot R)u^{\prime }-(u\cdot u^{\prime })R]  \nonumber \\
&=&-[(\widehat{u}\cdot \widehat{R})\widehat{u}^{\prime }-(\widehat{u}\cdot 
\widehat{u}^{\prime })\widehat{R}]  \nonumber \\
&=&-\widehat{u}^{\prime }\cosh \alpha +\widehat{R}\sinh \alpha  \label{e5a}
\end{eqnarray}

\begin{equation}
|\widehat{f}|=1  \label{e5b}
\end{equation}
Where $\alpha $ refers to the angle between $u$ and $R$, $R\perp u^{\prime
},r=|R|,$ $\widehat{u}=u/ic,\widehat{u}^{\prime }=u^{\prime }/ic,\widehat{R}%
=R/r$. Suppose that the magnitude of the force $f$ has the classical form

\begin{equation}
|f|=k\frac{qq^{\prime }}{r^2}  \label{e6}
\end{equation}
Combination of Eq.(\ref{e6}) with (\ref{e4}), we obtain a modified Coulomb's
force

\begin{eqnarray}
f &=&\frac{kqq^{\prime }}{c^2r^3}[(u\cdot X)u^{\prime }-(u\cdot u^{\prime })X%
]  \nonumber \\
&=&\frac{kqq^{\prime }}{c^2r^3}[(u\cdot R)u^{\prime }-(u\cdot u^{\prime })R]
\label{e7}
\end{eqnarray}
This force is in the form of Lorentz force for the two particles.

It is follows from Eq.(\ref{e7}) that the force can be rewritten in terms of
4-vector components as

\begin{eqnarray}
f_\mu &=&qF_{\mu \nu }u_\nu  \label{e8a} \\
F_{\mu \nu } &=&\partial _\mu A_\nu -\partial _\nu A_\mu  \label{e8b} \\
A_\mu &=&\frac{kq^{\prime }}{c^2}\frac{u_\mu ^{\prime }}r  \label{e8c}
\end{eqnarray}
Where we have used the relations

\begin{equation}
\partial _\mu \left( \frac 1r\right) =-\frac {R_\mu }{r^3}  \label{e9}
\end{equation}

\subsection{Lorentz gauge condition}

From Eq.(\ref{e8c}), because of $u^{\prime }\perp R$ , we have

\begin{equation}
\partial _\mu A_\mu =\frac{kq^{\prime }u_\mu ^{\prime }}{c^2}\partial _\mu
\left( \frac 1r\right) =-\frac{kq^{\prime }u_\mu ^{\prime }}{c^2}\left( 
\frac{R_\mu }{r^3}\right) =0  \label{e11}
\end{equation}
It is known as the Lorentz gauge condition.

\subsection{Maxwell's equations}

To note that $R$ has three degrees of freedom on the condition $R\perp
u^{\prime }$, so we have

\begin{equation}
\partial _\mu R_\mu =3  \label{e12}
\end{equation}

\begin{equation}
\partial _\mu \partial _\mu \left( \frac 1r\right) =-4\pi \delta (R)
\label{e13}
\end{equation}
From Eq.(\ref{e8b}), we have

\begin{eqnarray}
\partial _\nu F_{\mu \nu } &=&\partial _\nu \partial _\mu A_\nu -\partial
_\nu \partial _\nu A_\mu =-\partial _\nu \partial _\nu A_\mu  \nonumber \\
&=&-\frac{kq^{\prime }u_\mu ^{\prime }}{c^2}\partial _\nu \partial _\nu
\left( \frac 1r\right) =\frac{kq^{\prime }u_\mu ^{\prime }}{c^2}4\pi \delta
(R)  \nonumber \\
&=&\mu _0J_\nu ^{\prime }  \label{e14}
\end{eqnarray}
where we define $J_\nu ^{\prime }=q^{\prime }u_\nu ^{\prime }\delta (R)$.
From Eq.(\ref{e8b}), by exchanging the indices and taking the summation of
them, we have

\begin{equation}
\partial _\lambda F_{\mu \nu }+\partial _\mu F_{\nu \lambda }+\partial _\nu
F_{\lambda \mu }=0  \label{e15}
\end{equation}
The Eq.(\ref{e14}) and (\ref{e15}) are known as the Maxwell's equations. For
continuous media, they are valid as well.

\subsection{Lienard-Wiechert potential}

From Maxwell's equations, we know there is a retardation time for action to
propagate between the two particles, let $d$ denote the distance from $%
q^{\prime }$ to $O$ in Fig.\ref{sfig1}, the retardation effect is measured by

\begin{equation}
r=c\Delta t=c\frac d{ic}=c\frac{\widehat{u}^{\prime }\cdot X}{ic}=\frac{%
u_\nu ^{\prime }(x_\nu ^{\prime }-x_\nu )}c  \label{e17}
\end{equation}
Then

\begin{equation}
A_\mu =\frac{kq^{\prime }}{c^2}\frac{u_\mu ^{\prime }}r=\frac{kq^{\prime }}c%
\frac{u_\mu ^{\prime }}{u_\nu ^{\prime }(x_\nu ^{\prime }-x_\nu )}
\label{e18}
\end{equation}

Obviously, Eq.(\ref{e18}) is known as the Lienard-Wiechert potential for a
moving particle.

\section{Attraction between electron pair}

In the preceding section we have devoted into electrodynamic subjects in
details, the purpose is to establish full confidence in Eq.(\ref{e7}),
re-given by

\begin{eqnarray}
f &=&\frac{kqq^{\prime }}{c^2r^3}[(u\cdot X)u^{\prime }-(u\cdot u^{\prime })X%
]  \nonumber \\
&=&\frac{kqq^{\prime }}{c^2r^3}[(u\cdot R)u^{\prime }-(u\cdot u^{\prime })R]
\label{e19}
\end{eqnarray}
from which we will in this section discuss a mechanism for superconductivity.

Obviously, for an electron pair, the first term of Eq.(\ref{e19}) can give
an attraction between the two electrons in certain situations, the second
term represents a repulsion which contributes to classical Coulomb's force.
To note that this attraction of an electron pair requires no phonon
exchange, the attraction is definitely distinguishable from that in Cooper
pair of BCS theory.

In superconductors, there are, probably at least, two kinds of collective
motions which can eliminate the repulsion between two electrons and let the
attraction being dominant , one is the combination of lattice and electron
gas, another is electron gas themselves.

In traditional superconductor, the transition temperature are fairly low,
the electron gas and lattice must combine to depress the repulsion in
electron pair, because the isotrope effect makes it clear that ions in the
metal play an essential role in superconductivity. By this view point, we
can arrive at the same consequences as BCS theory.

In high $T_c$ superconductor, for certain situations, the lattice may play
less-important role in eliminating the repulsion in electron pair, the
repulsion may be removed by the electron gas itself.

In normal material without superconductivity, there must be in balance
between the repulsion and attraction in an electron pair, so that the fermi
gas theory plays good enough for explaining metallic properties, where the
electrons can be regarded as free electrons.

In the other hand, in some material, it is possible that the repulsion of
electron pair becomes dominant, thus the material will have opposite
behavior with respect to the superconductivity at low temperatures.

\section{Quantum wave equations}

No doubt, we honestly believe that Pythagoras theorem is valid in every
point in an inertial frame of reference, this requirement is an abstract
constraint on motion behavior of particle in the space-time.

Consider a particle displacing $(dx_1,dx_2,dx_3)$ in time interval $dt$ at
speed $v$, Pythagoras theorem is written in the form:

\begin{equation}
(dx_1)^2+(dx_2)^2+(dx_3)^2=(vdt)^2  \label{a1}
\end{equation}
Using the above equation, we find

\begin{eqnarray}
&&(dx_1)^2+(dx_2)^2+(dx_3)^2-(cdt)^2  \nonumber \\
&=&(vdt)^2-(cdt)^2  \nonumber \\
&=&(icdt)^2[1-(\frac vc)^2]  \label{a2}
\end{eqnarray}
With the help of new notations $x_4=ict$ and $d\tau =dt\sqrt{1-v^2/c^2}$ ,
Defining 4-vector velocity

\begin{eqnarray}
u_1 &=&\frac{dx_1/dt}{\sqrt{1-v^2/c^2}}=\frac{dx_1}{d\tau }  \label{a3} \\
u_2 &=&\frac{dx_2/dt}{\sqrt{1-v^2/c^2}}=\frac{dx_2}{d\tau }  \label{a31} \\
u_3 &=&\frac{dx_3/dt}{\sqrt{1-v^2/c^2}}=\frac{dx_3}{d\tau }  \label{a4} \\
u_4 &=&\frac{ic}{\sqrt{1-v^2/c^2}}=\frac{dx_4}{d\tau }  \label{a41}
\end{eqnarray}
Eq.(\ref{a2}) can be rewritten as

\begin{equation}
u_1^2+u_2^2+u_3^2+u_4^2=u_\mu u_\mu =-c^2  \label{a5}
\end{equation}
Where Greek index $\mu $ takes on 1,2,3 and 4. Multiplying the last equation
with the rest mass $m$ of the particle, we obtain

\begin{equation}
\frac{d(mu_\mu u_\mu )}{d\tau }=2m\mathbf{u\cdot }\frac{d\mathbf{u}}{d\tau }%
+2mu_4\frac{du_4}{d\tau }=0  \label{a6}
\end{equation}
Separating the last equation into two equations by using new notations $%
\mathbf{f}$ and $f_4$, we have

\begin{eqnarray}
m\frac{d\mathbf{u}}{d\tau } &=&\mathbf{f}  \label{a7} \\
m\frac{du_4}{d\tau } &=&-\frac{\mathbf{u\cdot f}}{u_4}=f_4  \label{a8}
\end{eqnarray}
The above equations represents the well-known relativistic dynamics given by

\begin{equation}
m\frac{du_\mu }{d\tau }=f_\mu  \label{a9}
\end{equation}
You see, indeed, the relativistic mechanics can be derived from ancient
Pythagoras theorem. For the details, please see the paper\cite{Cui1}

In an electromagnetic field, the dynamic equation [Eq.(\ref{a9})] is valid
at every point in the space-time, no mater whether there is an actual
particle passing the point considered, this means that there is a 4-vector
velocity $u$ at the point regardless of whether there exists a particle, in
other words, the 4-vector velocity is the geometric character of the point
in the electromagnetic field, it reflects some requirement arisen from
Pythagoras theorem, Every 4-vector velocity at every point forms a 4-vector
velocity field $u(x_1,x_2,x_3,x_4)$ in the space-time ( like the
geometrization of gravitational field ). The right side of Eq.(\ref{a9}) is
electromagnetic field, while the left side of Eq.(\ref{a9}) is 4-vector
velocity field, apple=apple, banana=banana, field=field, we have

\begin{eqnarray}
f_\mu &=&qF_{\mu \nu }u_\nu =qu_\nu (\partial _\mu A_\nu -\partial _\nu
A_\mu )  \label{a10} \\
m\frac{du_\mu }{d\tau } &=&m\frac{dx_\nu }{d\tau }(\frac{\partial u_\mu }{%
\partial x_\nu })=u_\nu \partial _\nu (mu_\mu )  \label{a11}
\end{eqnarray}
Substituting them back into Eq.(\ref{a9}), and re-arranging their terms, we
obtain

\begin{eqnarray}
u_\nu \partial _\nu (mu_\mu +qA_\mu ) &=&u_\nu \partial _\mu (qA_\nu ) 
\nonumber \\
&=&u_\nu \partial _\mu (mu_\nu +qA_\nu )-u_\nu \partial _\mu (mu_\nu ) 
\nonumber \\
&=&u_\nu \partial _\mu (mu_\nu +qA_\nu )-\frac 12\partial _\mu (mu_\nu u_\nu
)  \nonumber \\
&=&u_\nu \partial _\mu (mu_\nu +qA_\nu )-\frac 12\partial _\mu (-mc^2) 
\nonumber \\
&=&u_\nu \partial _\mu (mu_\nu +qA_\nu )  \label{a12}
\end{eqnarray}
Because the variables $\partial _\mu u_\nu $, $\partial _\mu A_\nu $, $%
\partial _\nu u_\mu $ and $\partial _\nu A_\mu $ are independent from $u_\nu 
$, a solution satisfying Eq.(\ref{a12}) is

\begin{equation}
\partial _\mu (mu_\nu +qA_\nu )=\partial _\nu (mu_\mu +qA_\mu )  \label{a13}
\end{equation}
According to Green's formula ( or Stokes's theorem ), the above equation
allows us to introduce a potential function $\Phi $ in mathematics, further
set $\Phi =-i\hbar \ln \psi $, we obtain a very important equation

\begin{equation}
(mu_\mu +qA_\mu )\psi =-i\hbar \partial _\mu \psi  \label{a14}
\end{equation}
where $\psi $ representing wave nature may be a complex mathematical
function, its physical meanings can be determined from experiments after the
introduction of the Planck's constant $\hbar $.

Substituting the last equation into Eq.(\ref{a5}) and eliminating $u_\mu $,
under different approximations we can derived out Klein-Gordon wave
equation, Dirac wave equation and Schrodinger wave equation, for the details
of the derivations please see the paper\cite{Cui2}.

From Eq.(\ref{a5}) and Eq.(\ref{a14}), we obtain a new quantum wave equation

\begin{equation}
-m^2c^2\psi ^2=(-i\hbar \partial _\mu -qA_\mu )\psi (-i\hbar \partial _\mu
-qA_\mu )\psi  \label{a15}
\end{equation}
Or in Gaussian units it is written as

\begin{equation}
-m^2c^2\psi ^2=(-i\hbar \partial _\mu -\frac qcA_\mu )\psi (-i\hbar \partial
_\mu -\frac qcA_\mu )\psi  \label{a16}
\end{equation}
Its precision is guaranteed by Pythagoras theorem. Thus, in the present
paper, we expect to find out more new results beyond Dirac, Klein-Gordon or
Schrodinger equations. In the following section, we will discuss spin in
details which plays an important role in superconducting states.

\section{Spin in atom}

\subsection{the electron in hydrogen atom}

In this section, we use Gaussian units, and use $m_e$ to denote the rest
mass of electron. We limit ourself to hydrogen atom and its spin.

In a spherical polar coordinate system $(r,\theta ,\varphi ,ict)$, the
nucleus of hydrogen atom provides a spherically symmetric potential $%
V(r)=e/r $ for the electron motion. The wave equation (\ref{a16}) for the
hydrogen atom in energy eigenstate $\psi (r,\theta ,\varphi )e^{iEt/\hbar }$
may be written in the spherical coordinates:

\begin{eqnarray}
\frac{m_e^2c^2}{\hbar ^2}\psi ^2 &=&(\frac{\partial \psi }{\partial r}%
)^2+(\frac 1r\frac{\partial \psi }{\partial \theta })^2+(\frac 1{r\sin
\theta }\frac{\partial \psi }{\partial \varphi })^2  \nonumber \\
&&+\frac 1{\hbar ^2c^2}(-E+\frac{e^2}r)^2\psi ^2  \label{b1}
\end{eqnarray}
By substituting $\psi =R(r)X(\theta )\phi (\varphi )$, we separate the above
equation into

\begin{eqnarray}
(\frac{\partial \phi }{\partial \varphi })^2+\kappa \phi ^2 &=&0  \nonumber
\\
&&  \label{b2} \\
(\frac{\partial X}{\partial \theta })^2+[\lambda -\frac \kappa {\sin
^2\theta }]X^2 &=&0  \nonumber \\
&&  \label{b3} \\
(\frac{\partial R}{\partial r})^2+[\frac 1{\hbar ^2c^2}(-E+\frac{e^2}r)^2-%
\frac{m_e^2c^2}{\hbar ^2}-\frac \lambda {r^2}]R^2 &=&0  \nonumber \\
&&  \label{b4}
\end{eqnarray}
The Eq.(\ref{b2}) can be solved immediately, with the requirement that $\phi
(\varphi )$ must be a periodic function, we find that its solution is given
by

\begin{eqnarray}
\phi &=&C_1e^{\pm i\sqrt{\kappa }\varphi }=C_1e^{im\varphi },\qquad
\label{b5} \\
m &=&\pm \sqrt{\kappa }=0,\pm 1,\pm 2,...  \nonumber
\end{eqnarray}
where $C_1$ is an integral constant.

Factoring Eq.(\ref{b3}), we get its two branches

\begin{equation}
\frac{\partial X}{\partial \theta }\pm iX\sqrt{\lambda -\frac{m^2}{\sin
^2\theta }}=0  \label{b6}
\end{equation}
It is easy to find their solutions

\begin{equation}
X(\theta )=C_2e^{\mp i\int \sqrt{\lambda -\frac{m^2}{\sin ^2\theta }}d\theta
}  \label{b7}
\end{equation}
where $C_2$ is an integral constant. The requirement of periodic function
for $X$ demands

\begin{equation}
\int\nolimits_0^{2\pi }\sqrt{\lambda -\frac{m^2}{\sin ^2\theta }}d\theta
=\pm 2\pi k\qquad k=0,1,2,...  \label{b8}
\end{equation}

Factoring Eq.(\ref{b2}), we get its two branches

\begin{equation}
\frac{\partial R}{\partial r}\pm \frac{iR}{\hbar c}\sqrt{(-E+\frac{e^2}%
r)^2-m_e^2c^4-\frac{\lambda \hbar ^2c^2}{r^2}}=0  \label{b9}
\end{equation}
and their solutions

\begin{equation}
R(r)=C_3e^{\mp \frac i{\hbar c}\int \sqrt{(-E+\frac{e^2}r)^2-m_e^2c^4-\frac{%
\lambda \hbar ^2c^2}{r^2}}dr}  \label{b10}
\end{equation}
where $C_3$ is an integral constant. The requirement that the radical wave
function forms a ''standing wave'' in the range from $r=0$ to $r=\infty $
demands

\begin{eqnarray}
\frac 1{\hbar c}\int\nolimits_0^\infty \sqrt{(-E+\frac{e^2}r)^2-m_e^2c^4-%
\frac{\lambda \hbar ^2c^2}{r^2}}dr &=&\pm \pi s  \label{b11} \\
\qquad s &=&0,1,2,...  \nonumber
\end{eqnarray}

Evaluating the definite integrals of Eq.(\ref{b8}) and Eq.(\ref{b11}) are
standard excises for contour integrals\cite{Marsden} in complex space.

\begin{figure}[htb]
\includegraphics[bb=210 325 380 665,clip]{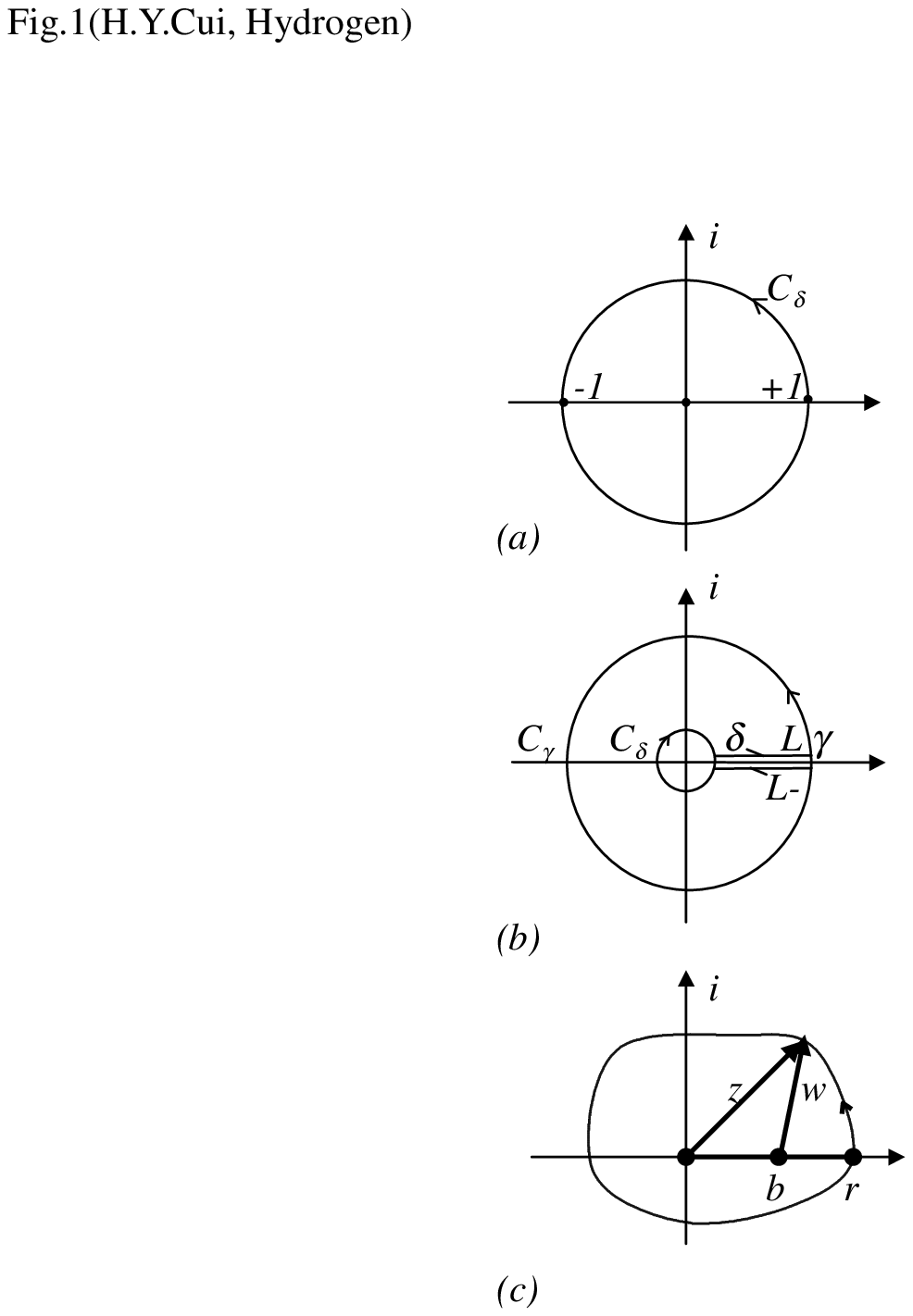}
\caption{Contours for evaluating integrals.}
\label{sfig2}
\end{figure}

Consider a contour $C_\delta $ which is a unit circle around zero, as shown
in Fig.\ref{sfig2}(a), using $z=e^{i\theta }$, we have

\begin{eqnarray}
I_1 &=&\int\nolimits_0^{2\pi }\sqrt{\lambda -\frac{m^2}{\sin ^2\theta }}%
d\theta =\int\nolimits_{C_\delta }\sqrt{\lambda +\frac{4m^2z^2}{(z^2-1)^2}}%
\frac{dz}{iz}  \nonumber \\
&=&\int\nolimits_{C_\delta }\frac{\sqrt{\lambda (z^2-1)^2+4m^2z^2}}{-(z^2-1)}%
\frac{dz}{iz}  \nonumber \\
&=&\int\nolimits_{C_\delta }(\frac 1z-\frac{1/2}{z-1}-\frac{1/2}{z+1})\sqrt{%
\lambda (z^2-1)^2+4m^2z^2}\frac{dz}i  \nonumber \\
&=&\int\nolimits_{C_\delta }\frac 1z\sqrt{\lambda (z^2-1)^2+4m^2z^2}\frac{dz}%
i  \nonumber \\
&&-\lim_{\eta \rightarrow 1}\int\nolimits_{C_\delta }\frac 1{2(z-\eta )}%
\sqrt{\lambda (z^2-1)^2+4m^2z^2}\frac{dz}i  \nonumber \\
&&-\lim_{\xi \rightarrow 1}\int\nolimits_{C_\delta }\frac 1{2(z+\xi )}\sqrt{%
\lambda (z^2-1)^2+4m^2z^2}\frac{dz}i  \nonumber \\
&=&\int\nolimits_{C_\delta }\frac{\sqrt{\lambda }+O(z^2)}z\frac{dz}i 
\nonumber \\
&&-\lim_{\eta \rightarrow 1}\int\nolimits_{C_\delta }\frac{|m|+O(z^2-1)}{%
z-\eta }\frac{dz}i  \nonumber \\
&&-\lim_{\xi \rightarrow 1}\int\nolimits_{C_\delta }\frac{|m|+O(z^2-1)}{%
z+\xi }\frac{dz}i  \nonumber \\
&=& 
\begin{array}{l}
\lceil \\ 
\langle \\ 
\lfloor
\end{array}
\begin{array}{l}
2\pi \sqrt{\lambda } \\ 
2\pi (\sqrt{\lambda }-|m|) \\ 
2\pi (\sqrt{\lambda }-2|m|)
\end{array}
\qquad 
\begin{array}{l}
|\eta |>1,|\xi |>1 \\ 
|\eta |<1or|\xi |<1 \\ 
|\eta |<1,|\xi |<1
\end{array}
\label{c0}
\end{eqnarray}
Where the integrand has the poles at $z=0$ and $z=\pm 1$, and we have chosen 
$\sqrt{(z^2-1)^2}=-(z^2-1)$. Comparing with Eq.(\ref{b8}) , the right side
of Eq.(\ref{b8}) is required to take plus sign, we obtain

\begin{equation}
\sqrt{\lambda }= 
\begin{array}{l}
\lceil \\ 
\langle \\ 
\lfloor
\end{array}
\begin{array}{l}
k \\ 
k+|m| \\ 
k+2|m|
\end{array}
\qquad 
\begin{array}{l}
|\eta |>1,|\xi |>1 \\ 
|\eta |<1or|\xi |<1 \\ 
|\eta |<1,|\xi |<1
\end{array}
\label{c1}
\end{equation}
We rename the integer $\lambda $ as $j^2$ for a convenient in the following,
i.e. $\lambda =j^2$.

Consider a contour $C$ , consisting of $C_\gamma $, $L_{-}$, $C_\delta $ and 
$L$ around zero in the plane as shown in Fig.\ref{sfig2}(b), the radius of
circle $C_\gamma $ is large enough and the radius of circle $C_\delta $ is
small enough. The integrand of the following equation has no pole inside the
contour $C$, so that we have

\begin{eqnarray}
&&\int\nolimits_C\sqrt{(-E+\frac{e^2}z)^2-m_e^2c^4-\frac{j^2\hbar ^2c^2}{z^2}%
}dz  \nonumber \\
&=&\int\nolimits_{C_\gamma }+\int\nolimits_{L_{-}}+\int\nolimits_{C_\delta
}+\int\nolimits_L=0  \label{c2}
\end{eqnarray}
The integrals in Eq.(\ref{c2}) are

\begin{eqnarray}
\int\nolimits_{C_\gamma } &=&\int\nolimits_{C_\gamma }\sqrt{(-E+\frac{e^2}%
z)^2-m_e^2c^4-\frac{j^2\hbar ^2c^2}{z^2}}dz  \nonumber \\
&=&\int\nolimits_{C_\gamma }[\sqrt{E^2-m_e^2c^4}-\frac{Ee^2}{\sqrt{%
E^2-m_e^2c^4}}\frac 1z+O(\frac 1{z^2})]dz  \nonumber \\
&=&\frac{-i2\pi Ee^2}{\sqrt{E^2-m_e^2c^4}}=\frac{2\pi Ee^2}{\sqrt{%
m_e^2c^4-E^2}}  \label{c3}
\end{eqnarray}

\begin{eqnarray}
\int\nolimits_{C_\delta } &=&\int\nolimits_{C_\delta }\sqrt{(-E+\frac{e^2}%
z)^2-m_e^2c^4-\frac{j^2\hbar ^2c^2}{z^2}}dz  \nonumber \\
&=&\int\nolimits_{C_\delta }\frac{\sqrt{(-Ez+e^2)^2-m_e^2c^4z^2-j^2\hbar
^2c^2}}zdz  \nonumber \\
&=&\int\nolimits_{C_\delta }\frac{\sqrt{e^4-j^2\hbar ^2c^2}+O(z)}zdz 
\nonumber \\
&=&-i2\pi \sqrt{e^4-j^2\hbar ^2c^2}=-2\pi \sqrt{j^2\hbar ^2c^2-e^4}
\label{c4}
\end{eqnarray}
Because the integrand is a multiple-valued function, when the integral takes
over the path $L_{-}$ we have $z=e^{i2\pi }re^{+0i}$, thus

\begin{equation}
\int\nolimits_{L-}=\int\nolimits_\gamma ^\delta \sqrt{e^{-i2\pi }(...)}%
=-\int\nolimits_\gamma ^\delta =\int\nolimits_\delta ^\gamma =\int\nolimits_L
\label{c50}
\end{equation}
For a further manifestation, to define $z-b=w=\rho e^{i\beta }$, where

\begin{equation}
b=\frac{E^2+e^4-m_e^2c^4r^2-j^2\hbar ^2c^2}{2Ee^2}  \label{c51}
\end{equation}
We have

\begin{eqnarray}
\int\nolimits_{L-} &=&\int\nolimits_{L-}\frac{\sqrt{-2Ee^2}\sqrt{z-b}}zdz 
\nonumber \\
&=&\int\nolimits_{L-}\frac{\sqrt{-2Ee^2}\sqrt{w}}zdz  \nonumber \\
&=&\int\nolimits_{L-}\frac{\sqrt{-2Ee^2\rho }e^{i\beta /2}}zdz  \nonumber \\
&=&\int\nolimits_{L(\gamma \rightarrow \delta )}\frac{\sqrt{-2Ee^2\rho }%
e^{i(\beta -2\pi )/2}}{ze^{-i2\pi }}dz  \nonumber \\
&=&-\int\nolimits_{L(\gamma \rightarrow \delta )}\frac{\sqrt{-2Ee^2\rho }%
e^{i\beta /2}}zdz  \nonumber \\
&=&-\int\nolimits_\gamma ^\delta =\int\nolimits_\delta ^\gamma
=\int\nolimits_L  \label{c52}
\end{eqnarray}
The relation of $z$ and $w$ has shown in Fig.\ref{sfig2}(c), to note that $w$
rotates around zero with $z$.

\begin{eqnarray}
\int\nolimits_L &=&\frac 12(\int\nolimits_L+\int\nolimits_{L-})=-\frac
12(\int\nolimits_{C_\gamma }+\int\nolimits_{C_\delta })  \nonumber \\
&=&-\frac{\pi Ee^2}{\sqrt{m_e^2c^4-E^2}}+\pi \sqrt{j^2\hbar ^2c^2-e^4}
\label{c5}
\end{eqnarray}
Thus Eq.(\ref{b11}) becomes

\begin{eqnarray}
\pm \pi s &=&\frac 1{\hbar c}\int\nolimits_0^\infty \sqrt{(-E+\frac{e^2}%
z)^2-m_e^2c^4-\frac{j^2\hbar ^2c^2}{z^2}}dr  \nonumber \\
&=&-\frac{\pi E\alpha }{\sqrt{m_e^2c^4-E^2}}+\pi \sqrt{j^2-\alpha ^2}
\label{c6}
\end{eqnarray}
where $\alpha =e^2/\hbar c$ is known as the fine structure constant. The
left side of the last equation is required to take minus, then we obtain the
positive energy levels given by

\begin{equation}
E=m_ec^2\left[ 1+\frac{\alpha ^2}{(\sqrt{j^2-\alpha ^2}+s)^2}\right]
^{-\frac 12}  \label{c7}
\end{equation}

According to Eq.(\ref{c7}), we find that the first branch and third branch
in Eq.(\ref{c1}) are degenerated into the second branch for a fixed $j$. So
we need only to define $j=k+|m|$. Because $k\geq 0$ , and $j\neq 0$ in Eq.(%
\ref{c7}), we find $j=1,2,3...$.

The result, Eq.(\ref{c7}), is completely the same as the calculation of
Dirac wave equation\cite{Schiff}, it is just the fine structure of hydrogen
atom.

\subsection{hydrogen atom in an uniform magnetic field}

If we put the hydrogen atom into an external uniform magnetic field $B$
which is along $z$ axis with vector potential $(A_r,A_\theta ,A_\varphi
)=(0,0,\frac 12r\sin \theta B)$, then according to Eq.(\ref{a16}) the wave
equation is given by

\begin{eqnarray}
\frac{m_e^2c^2}{\hbar ^2}\psi ^2 &=&(\frac{\partial \psi }{\partial r}%
)^2+(\frac 1r\frac{\partial \psi }{\partial \theta })^2+\frac 1{\hbar
^2c^2}(-E+\frac{e^2}r)^2\psi ^2  \nonumber \\
&&+(\frac 1{r\sin \theta }\frac{\partial \psi }{\partial \varphi }-\frac{%
er\sin \theta B}{i2c\hbar }\psi )^2  \label{d1}
\end{eqnarray}
By substituting $\psi =R(r)X(\theta )\phi (\varphi )$, we separate the above
equation into

\begin{eqnarray}
\frac{\partial \phi }{\partial \varphi }-\kappa \phi &=&0  \nonumber \\
&&  \label{d2} \\
(\frac{\partial X}{\partial \theta })^2+[(\frac \kappa {\sin \theta }-\frac{%
e\sin \theta r^2B}{i2c\hbar })^2+\xi (r)]X^2 &=&0  \nonumber \\
&&  \label{d3} \\
(\frac{\partial R}{\partial r})^2+[\frac 1{\hbar ^2c^2}(-E+\frac{e^2}r)^2-%
\frac{m_e^2c^2}{\hbar ^2}-\frac{\xi (r)}{r^2}]R^2 &=&0  \nonumber \\
&&  \label{d4}
\end{eqnarray}
Where we have used the unknown constant $\kappa $ and function $\xi (r)$ to
connect these separated equations. Eq.(\ref{d2}) has the solution

\begin{equation}
\phi =C_1e^{im\varphi },\qquad \kappa =im,\qquad m=0,\pm 1,\pm 2,...
\label{d5}
\end{equation}
Expanding Eq.(\ref{d3}) and neglecting the term $O(B^2)$, we have a term $-%
\frac{mer^2B}{c\hbar }$ in it, by moving this term into Eq.(\ref{d4})
through $\xi (r)=\lambda +\frac{mer^2B}{c\hbar }$, we obtain

\begin{eqnarray}
(\frac{\partial X}{\partial \theta })^2+[\lambda -\frac{m^2}{\sin ^2\theta }%
]X^2 &=&0  \nonumber \\
&&  \label{d7} \\
(\frac{\partial R}{\partial r})^2+[\frac 1{\hbar ^2c^2}(-E+\frac{e^2}r)^2-%
\frac{m_e^2c^2}{\hbar ^2}-\frac{meB}{c\hbar }-\frac \lambda {r^2}]R^2 &=&0 
\nonumber \\
&&  \label{d8}
\end{eqnarray}
The above two equations are the same as Eq.(\ref{b3}) and Eq.(\ref{b4}),
except for the additional constant term $-meB/c\hbar $. After the similar
calculation in the preceding section, we obtain the energy levels of
hydrogen atom in the magnetic field given by

\begin{equation}
E=\sqrt{m_e^2c^4+mec\hbar B}\left[ 1+\frac{\alpha ^2}{(\sqrt{j^2-\alpha ^2}%
+s)^2}\right] ^{-\frac 12}  \label{d9}
\end{equation}

\subsection{spectroscopic notation}

In the usual spectroscopic notation of quantum mechanics, four quantum
numbers: $n$, $l$, $m_l$ and $m_s$ are used to specify the state of an
electron in an atom. After the comparison, we get the relations between the
usual notation and our notation.

\begin{eqnarray}
n &=&j+s,\quad s=0,1,...;j=1,2,....  \label{f1} \\
l &=&j-1,  \label{f2} \\
\quad \max (m_l) &=&\max (m)-1  \label{f3}
\end{eqnarray}
We find that $j$ takes over $1,2,...,n$; for a fixed $j$ (or $l$), $m$ takes
over $-(l+1),-l,...,0,...,l,l+1$. In the present work, spin quantum number
is absent.

\subsection{Zeeman splitting}

According to Eq.(\ref{d9}), for a fixed $(n,l)$, equivalent to $(n,j=l+1)$,
the energy level of hydrogen atom will split into $2l+3$ energy levels in
magnetic field, given by

\begin{equation}
E=(m_ec^2+\frac{me\hbar B}{2m_ec})\left[ 1+\frac{\alpha ^2}{(\sqrt{%
j^2-\alpha ^2}+s)^2}\right] ^{-\frac 12}+O(B^2)  \label{f4}
\end{equation}
Considering $m=-(l+1),-l,...,0,...,l,l+1$, this effect is equivalent to the
usual Zeeman splitting in the usual quantum mechanics given by

\begin{equation}
E=E_{nl}+\frac{(m_l\pm 1)e\hbar B}{2m_ec}  \label{f5}
\end{equation}
But our work works on it without spin concept.

\subsection{angular momentum, Stern-Gerlach experiment and spin}

From Eq.(\ref{a14}), the angular momentum of the electron in hydrogen atom
is given by

\begin{eqnarray}
\mathbf{J} &=&J_\varphi \mathbf{e}_\varphi +J_z\mathbf{e}_z  \nonumber \\
&=&rm_eu_\theta \mathbf{e}_\varphi +r\sin \theta m_eu_\varphi \mathbf{e}_z 
\nonumber \\
&=&\frac 1\psi (-i\hbar \frac{\partial \psi }{\partial \theta })\mathbf{e}%
_\varphi +\frac 1\psi (-i\hbar \frac{\partial \psi }{\partial \varphi })%
\mathbf{e}_z  \nonumber \\
&=&\hbar \sqrt{\lambda -\frac{m^2}{\sin ^2\theta }}\mathbf{e}_\varphi
+m\hbar \mathbf{e}_z  \nonumber \\
&=&\hbar \sqrt{(k+|m|)^2-\frac{m^2}{\sin ^2\theta }}\mathbf{e}_\varphi
+m\hbar \mathbf{e}_z  \label{f6}
\end{eqnarray}
According to Eq.(\ref{c7}), there should be three ground states ($j=1$).

\begin{eqnarray}
1 &:&k=0,m=1,\mathbf{J}=i(\cot \theta )\hbar \mathbf{e}_\varphi +\hbar 
\mathbf{e}_z  \label{f7} \\
2 &:&k=0,m=-1,\mathbf{J}=i(\cot \theta )\hbar \mathbf{e}_\varphi -\hbar 
\mathbf{e}_z  \label{f8} \\
3 &:&k=1,m=0,\mathbf{J}=\hbar \mathbf{e}_\varphi  \label{f9}
\end{eqnarray}

\begin{figure}[htb]
\includegraphics[bb=230 470 390 590,clip]{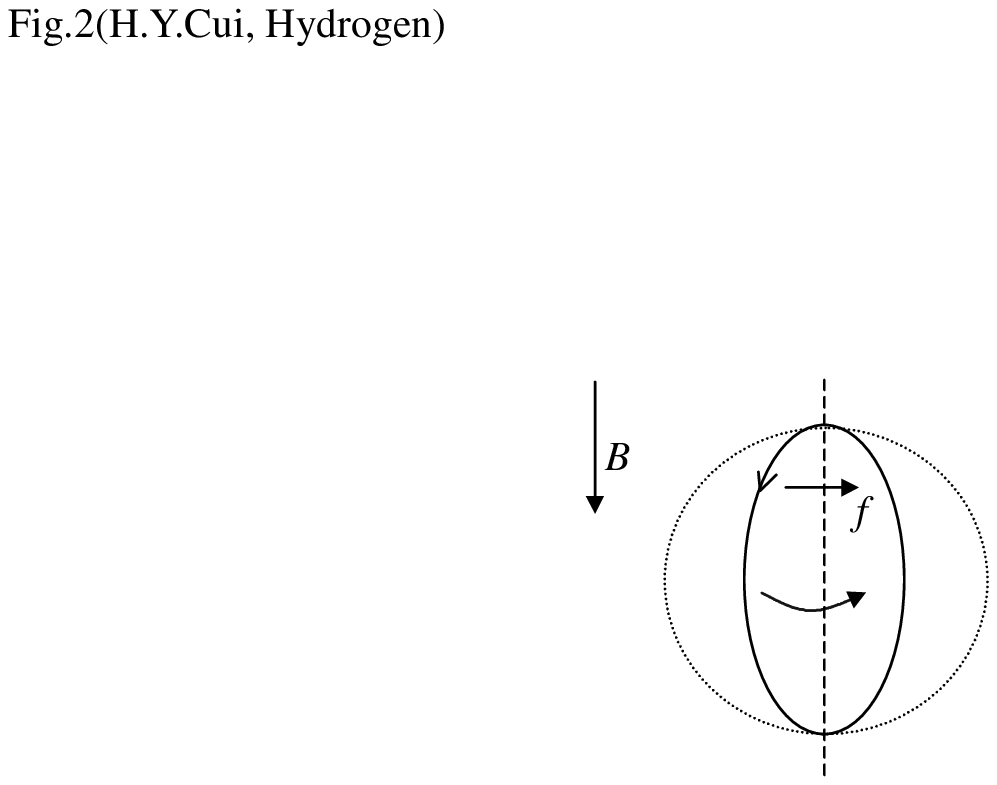}
\caption{The motion of the electron in hydrogen atom.}
\label{sfig3}
\end{figure}

Why did the hydrogen atoms split into two branches in Stern-Gerlach
experiment? To note that $J_\varphi \rightarrow \infty $ when $\theta
\rightarrow 0,\pi $ for the ground states1 and 2, maybe $J_\varphi $ locks
the alignment of the electrons with external magnetic fields along $z$-axis,
considering the directions of Lorentz forces acting on the currents as shown
in Fig.\ref{sfig3} schematically (despite it is imaginary), so that the
angular momenta $J_z\mathbf{=\pm \hbar }$ of the electrons can not be
subject to external magnetic direction as the usual way. Conversely, the
ground state 3 can rotates its orientation like a coil, disappeared within
one of the two branches in Stern-Gerlach experiment, In the fact, maybe the
ground state 3 has changed into the ground states 1 or 2 before it rotates
to its usual destination because of its increasing or decreasing $J_z$. This
explanation has drown an image for spin.

\subsection{the shell of atom}

In atom with many electrons, two electrons can form a pair with $%
\bigtriangleup J_z\mathbf{=\pm \hbar }$, the reason of pairing arises
probably from the alignment of their $J_\varphi $ and $J_z$, the situation
can imagined as an analogy to the combination of the ground states 1 and 2
of hydrogen atom.

\begin{figure}[htb]
\includegraphics[bb=210 525 420 630,clip]{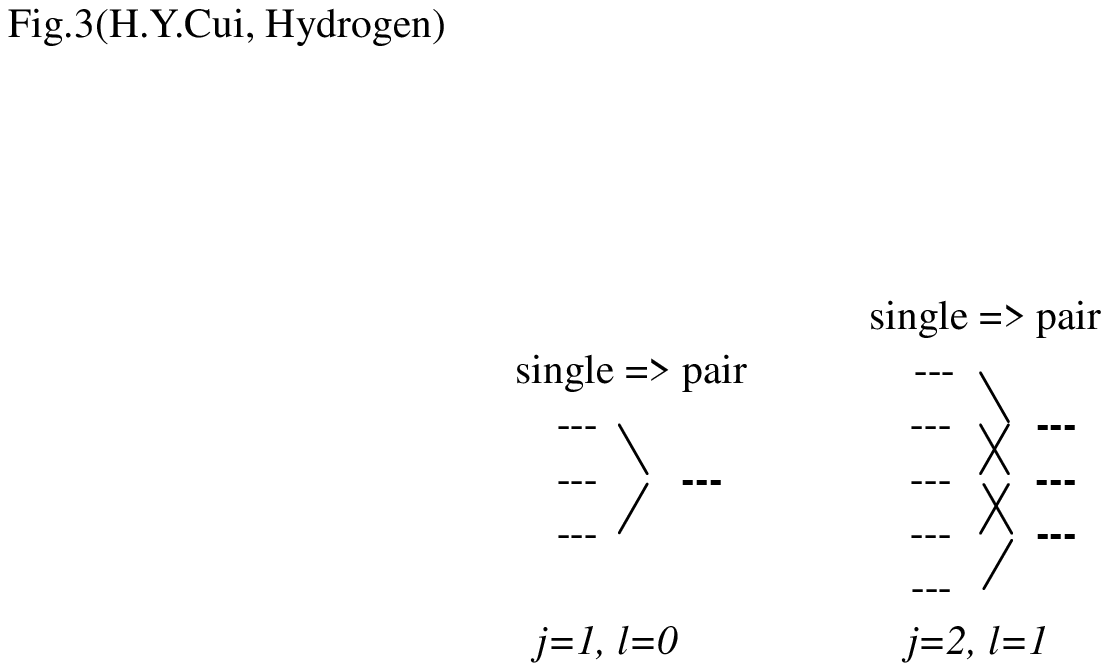}
\caption{Single-states merge into pair-states.}
\label{sfig4}
\end{figure}

For a fixed $l$ (or $j$), $m$ takes over $2l+3$ values, thus the electrons
with $2l+3$ single-states can only form $2l+1$ pair-states, as shown in Fig.%
\ref{sfig4}, the $l$-shell contains $2(2l+1)$ electrons.

Although the single state of $m=0$, like the ground state 3 of hydrogen
atom, maybe unstable, it in a pair can be stable because it can align its $%
J_\varphi $ with its counterpart electron.

Although the shell structure of atom based on our work is the same as the
usual one, the present work predicts that hydrogen atom has $2l+3$
degenerate states for a fixed $l$.

\subsection{an open problem}

In order to meet the fine structure of hydrogen atom, we have chosen plus
sign for $k$ in Eq.(\ref{b8}) and minus sign for $s$ in Eq.(\ref{b11}). In
other words, there are more states we have not discussed yet, which
correspond to minus sign for $k$ in Eq.(\ref{b8}) or plus sign for $s$ in
Eq.(\ref{b11}). Two situations must be considered, there probably exists an
unknown constraint forbidding the states, or experiment has not payed
attention to observe the states, this is an open question.

\section{Superconducting states}

Now we return to superconducting states. In the preceding section we have
devoted into hydrogen atom and spin subjects in details, the purpose is to
establish full confidence in Eq.(\ref{a14}), re-given by

\begin{equation}
(mu_\mu +qA_\mu )\psi =-i\hbar \partial _\mu \psi  \label{h1}
\end{equation}
In this section $m$ is the rest mass. The wave function can be rewritten in
integral form as

\begin{equation}
\psi =e^{\frac i\hbar \int_{x_0}^x(mu_\mu +qA_\mu )dx_\mu +i\theta }
\label{h2}
\end{equation}
Where $\theta $ is an integral constant, $x_0$ and $x$ are the initial point
and final point of the integral with arbitrary path in the space-time.

Let us take the integration of Eq.(\ref{h2}) over a closed path $L$ in the
space, at any instant $t(dx_4=0)$, the single-valued wave function of Eq.(%
\ref{h2}) requires

\begin{equation}
\frac 1\hbar \oint_L(m\mathbf{u}+q\mathbf{A)\cdot }d\mathbf{l=}2\pi b\qquad
b=0,\pm 1,\pm 2,...  \label{h3}
\end{equation}
Where $d\mathbf{l}$ is an element of the integral path.

\subsection{quantized magnetic flux in superconducting ring}

Let us take integral path $L$ in superconducting ring, because the
electronic current is zero in the interior, i.e. $\mathbf{u}=0$, thus the
magnetic flux through $L$ is given from \ref{h3} by

\begin{equation}
\phi =\oint_L\mathbf{B\cdot }d\mathbf{\sigma =}\oint_L\mathbf{A\cdot }d%
\mathbf{l=}\frac{2\pi \hbar b}q  \label{h4}
\end{equation}
Where $d\mathbf{\sigma }$ is an element of area on the surface bounded by
the integral path $L$. By experiment $q=2e$, the charge of an electron pair,
thus the flux through the ring is quantized in integral multiples of $\pi
\hbar b/e$.

\subsection{London equation}

Let $n$ denote the density of electrons in superconductor, the electronic
current is given by

\begin{equation}
\mathbf{j}=-ne\mathbf{u}  \label{h5}
\end{equation}
Then Eq.(\ref{h3}) can be rewritten as

\begin{eqnarray}
&&\frac{-m}{ne\hbar }\oint_L(\mathbf{j}+\frac{ne^2}m\mathbf{A)\cdot }d%
\mathbf{l}  \nonumber \\
&=&\frac{-m}{ne\hbar }\oint_L(\nabla \times \mathbf{j}+\frac{ne^2}m\mathbf{%
B)\cdot }d\mathbf{\sigma =}2\pi s  \label{h6}
\end{eqnarray}
If the closed integral path contains no vortex (or fluxoid), at where we
obtain

\begin{equation}
\nabla \times \mathbf{j}+\frac{ne^2}m\mathbf{B=0}  \label{h7}
\end{equation}
It is the well known London equation.

\subsection{Meissner effect}

\ According to a Maxwell equation

\begin{equation}
\nabla \times \mathbf{B}=\mu _0\mathbf{j}  \label{h8}
\end{equation}
from London equation we obtain

\begin{equation}
\nabla ^2\mathbf{B}=\mathbf{B}/\lambda _L^2  \label{h9}
\end{equation}
Where $\lambda _L^2=m/(\mu _0ne^2)$ is a constant called as the London
penetration depth. Near the surface of superconductor, the magnetic field
along the depth direction $z$ is given by

\begin{equation}
B(z)=B(0)\exp (-z/\lambda _L)  \label{h10}
\end{equation}
Thus for superconductor, the London equation leads to the Meissner effect.

\subsection{Josephson superconductor tunneling}

Before we discuss Josepheson effect, we firstly discuss Aharonov-Bohm effect.

\paragraph{Aharonov-Bohm effect}

\begin{figure}[ht]
\includegraphics[bb=110 540 310 720,clip]{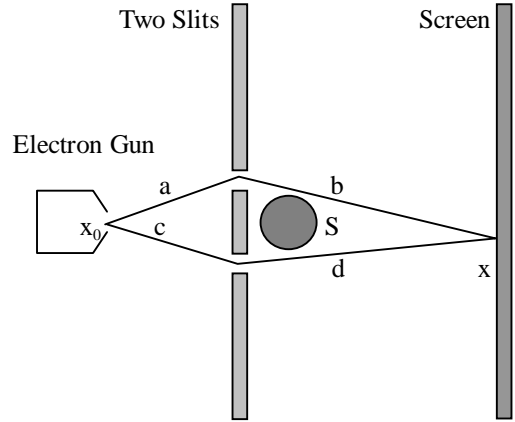}
\caption{A diffraction experiment with adding a solenoid.}
\label{sfig5}
\end{figure}

Let us consider the modification of two slit experiment, as shown in Fig.\ref
{sfig5}. Between the two slits there is located a tiny solenoid S, designed
so that a magnetic field perpendicular to the plane of the figure can be
produced in its interior. No magnetic field is allowed outside the solenoid,
and the walls of the solenoid are such that no electron can penetrate to the
interior. There are two paths $l_1$ and $l_2$ bypassing the solenoid from
the electron gun to the screen, the wave function $\psi $ is given by

\begin{equation}
\psi =e^{\frac i\hbar \int\nolimits_{x_0(l_1)}^x(mu_\mu +qA_\mu )dx_\mu
}+e^{\frac i\hbar \int\nolimits_{x_0(l_2)}^x(mu_\mu +qA_\mu )dx_\mu }
\label{p9}
\end{equation}
because $l_1$ and $l_2$ are equivalent for Eq.(\ref{h2}). The probability is
given by

\begin{eqnarray}
W &=&\psi (x)\psi ^{*}(x)  \nonumber \\
&=&2+e^{\frac i\hbar \int\nolimits_{x_0(l_1)}^x(mu_\mu +qA_\mu )dx_\mu
-\frac i\hbar \int\nolimits_{x_0(l_2)}^x(mu_\mu +qA_\mu )dx_\mu }  \nonumber
\\
&&+e^{\frac i\hbar \int\nolimits_{x_0(l_2)}^x(mu_\mu +qA_\mu )dx_\mu -\frac
i\hbar \int\nolimits_{x_0(l_1)}^x(mu_\mu +qA_\mu )dx_\mu }  \nonumber \\
&=&2+2\cos [\frac p\hbar (l_1-l_2)+\frac 1\hbar
\int\nolimits_{x_0(l_1)}^xqA_\mu dx_\mu  \nonumber \\
&&-\frac 1\hbar \int\nolimits_{x_0(l_2)}^xqA_\mu dx_\mu ]  \nonumber \\
&=&2+2\cos [\frac p\hbar (l_1-l_2)+\frac 1\hbar \oint_{(l_1+\overline{l_2}%
)}qA_\mu dx_\mu ]  \nonumber \\
&=&2+2\cos [\frac p\hbar (l_1-l_2)+\frac{q\phi }\hbar ]  \label{p10}
\end{eqnarray}
where $\overline{l_2}$ denotes the inverse path to the path $l_2$, $\phi $
is the magnetic flux that passes through the surface between the paths $l_1$
and $\overline{l_2}$, and it is just the flux inside the solenoid.

Now, constructive (or destructive) interference occurs when

\begin{equation}
\frac p\hbar (l_1-l_2)+\frac{q\phi }\hbar =2\pi b\quad (or\quad b+\frac 12)
\label{p11}
\end{equation}
where $b$ is an integer. We know that this effect is just the Aharonov-Bohm
effect which was shown experimentally in 1960.

\paragraph{Josephson effect}

Consider an insulator film occupying the space of rang $(0,\delta )$ in $x$%
-axis with an applied voltage $V$, according to Eq.(\ref{h2}), set the
origin at $x=0$, the wave function of electron pair in the region $x>\delta $
can be calculated by taking the integral path $l_1$ across the insulator, it
gives

\begin{equation}
\psi _1=e^{\frac i\hbar \int_0^\delta pdx+\frac i\hbar \int_\delta
^xpdx+\frac i\hbar \int_0^tqA_4dx_4+i\theta _1}  \label{h12}
\end{equation}
the wave function of electron pair in the region $x>\delta $ can also be
calculated by taking the integral path $l_2$ around the circuit without
crossing the insulator, it gives

\begin{equation}
\psi _2=e^{\frac i\hbar \int_\delta ^xpdx+i\theta _2}  \label{h13}
\end{equation}
To note that the electrons have the same momentum $p$ in the superconductors
whereas have an imaginary momentum in the insulator, to denote $%
\int_0^\delta pdx=i\hbar k\delta $, we obtain the wave function

\begin{eqnarray}
\psi  &=&\psi _1+\psi _2  \nonumber \\
&=&e^{-k\delta +\frac i\hbar \int_\delta ^xpdx+\frac{i2eVt}\hbar +i\theta
_1}+e^{\frac i\hbar \int_\delta ^xpdx+i\theta _2}  \label{h14}
\end{eqnarray}
According to Eq.(\ref{h1}), the superconducting current in $x>\delta $
region is given by

\begin{eqnarray}
j &=&nqu=\frac{nq}mp=-i\frac{q\hbar }{2m}(\psi ^{*}\frac{\partial \psi }{%
\partial x}-\psi \frac{\partial \psi ^{*}}{\partial x})  \nonumber \\
&=&\frac{2e}mpe^{-k\delta }\sin [\frac{2eVt}\hbar +i(\theta _1-\theta _2)]
\label{h15}
\end{eqnarray}
The result is just the Josephson tunneling effect, including DC Josephson
effect and AC Josephson effect.

\section{General superconducting states}

It is easy to find that the electrons in Aharonov-Bohm experiment can
regarded as being in a superconducting state because of lacking
lattice-scattering, the similar situations like electron two split
experiment etc. give us the same inspiration. Even the elections in an atom
can also be regarded as in a superconducting state. The feeling becomes much
strong when we recognize that the two electrons with up and down spins in an
atom always form a pair in the sell of the atom. Under the guidance of the
general superconducting states, we think that the electron pairs in atom
must have the same mechanism with Cooper pairs in superconductor. The
relativistic mechanism of superconductivity suggested in the present paper
just satisfies this requirement.

\section{Conclusion}

According to the theory of relativity, the relativistic Coulomb's force
between an electron pair is composed of two parts, the main part is
repulsive, while the rest part can be attractive in certain situations. The
relativistic attraction of an electron pair provides a insight into the
mechanism of superconductivity.

In superconductor, there are, probably at least, two kinds of collective
motions which can eliminate the repulsion between two electrons and let the
attraction being dominant, the first is the combination of lattice and
electron gas, accounting for  traditional superconductivity; the second is
the electron gas themselves, accounting for high $T_c$ superconductivity.

In usual materials, there is a good balance between the repulsion and
attraction of an electron pair, the electrons are regarded as free electrons
so that Fermi gas theory plays very well. But in some materials, when the
repulsion dominates electron pairs, the electron gas will has a behavior
opposite to superconductivity.

The superconducting states are discussed in terms of relativistic quantum
theory in the present paper, some significant results are obtained,
including quantized magnetic flux, London equation, Meissner effect and
Josephson effect.

\end{document}